\documentclass[a4paper,12pt]{article}

\usepackage[T1]{fontenc}
\usepackage{lmodern}
\usepackage[margin=25truemm]{geometry}
\usepackage{amsmath}
\usepackage{amssymb}
\usepackage{amsthm}
\usepackage{bm}
\usepackage{graphicx}
\usepackage{tabularx}
\usepackage{caption}
\usepackage{mathtools}
\usepackage{stmaryrd}
\usepackage[
  pdfencoding=auto,
  pdfborder={0 0 0},
  bookmarksnumbered=true
]{hyperref}

\newcommand{\eq}[1]{\begin{align}#1\end{align}}

\newcommand{\pd}{\partial}

\newcommand{\nn}{\nonumber}

\newcommand{\DD}{\mathsf D}
\newcommand{\MM}{\mathsf M}
\newcommand{\LL}{\mathsf L}
\newcommand{\TT}{\mathsf T}
\newcommand{\SSS}{\mathsf S}
\newcommand{\CC}{\mathsf C}
\newcommand{\llb}{\llbracket}
\newcommand{\rrb}{\rrbracket}

\DeclareMathOperator{\rank}{rank}
\DeclareMathOperator{\im}{im}

\begin{document}

\begin{center}
{\LARGE \bf
A linear-algebraic formulation of dimensional analysis with constraints
}

\vspace{9mm}

{\large Umpei Miyamoto}

\vspace{2mm}

Research and Education Center for Comprehensive Science,\\
Akita Prefectural University, Akita 015-0055, Japan

\vspace{2mm}

umpei@akita-pu.ac.jp
\end{center}

\vspace{4mm}

\begin{abstract}
Dimensional analysis, especially Buckingham's $\pi$ theorem, reduces the number of variables by rewriting a relation in terms of dimensionless quantities. When variables are tied by definitions, constitutive laws, or other constraints, however, eliminating variables in advance can be awkward. We formulate dimensional analysis with constraints as linear algebra in logarithmic variables. Dimensional transformations and constraints are represented by subspaces, the effective number of independent dimensionless quantities is characterized by their intersection, and a matrix representation yields a systematic redundancy elimination procedure. Examples from falling motion, drag force, and stock-market indicators illustrate the scope and limitations of the method.
\end{abstract}

\section{Introduction}

Dimensional analysis, and in particular Buckingham's $\pi$ theorem, is a basic tool for reducing the number of variables by rewriting relations among quantities as relations among dimensionless quantities~\cite{buckingham1914}. Its strength is that it can restrict in advance which combinations of variables can be essential, even when the detailed governing equations of the system are unknown. This idea has long played an important role in model experiments, similarity laws, scaling arguments, and experimental design.

As a familiar example, consider a simple pendulum. Suppose that the period $T$ depends only on the length $L$ of the string, the gravitational acceleration $g$, and the mass $m$ of the bob. Buckingham's $\pi$ theorem implies that $T\sqrt{g/L}$ is dimensionless and that the mass $m$ cannot enter the formula for the period. Therefore the period must have the form
\eq{
T=c\sqrt{\frac{L}{g}},
\label{eq:pendulum-period}
}
where $c$ is a dimensionless constant. Solving the equation of motion gives $c=2\pi$, but dimensional analysis already fixes the possible form of the relation.

In many real problems, however, variables are connected by additional relations such as definitions, constitutive laws, or implicit equations. In such cases, the naive procedure of eliminating variables first and then applying the $\pi$ theorem is not always transparent. When there are many constraints, or when the elimination itself is cumbersome, even the question of which dimensionless quantities should remain can become unclear.

The use of logarithmic variables to turn multiplicative structures into additive ones is standard in dimensional analysis and scaling theory~\cite{sedov1993,barenblatt1996}. In this representation, changes of units, dimensionless quantities, and constraints can all be described in terms of linear subspaces and their intersections. From this viewpoint, dimensional analysis with constraints should admit a natural linear-algebraic formulation.

The purpose of this paper is to formulate dimensional analysis for constrained systems as linear algebra in logarithmic variables. By representing dimensional relations and constraints as linear structures in the same space, we characterize the effective number of independent dimensionless quantities geometrically. We also show that the relations among candidate dimensionless quantities are encoded by a matrix, denoted by $C$, and give a redundancy elimination procedure based on elementary row operations. The aim is to provide a framework in which constraints can be kept until the end, rather than eliminated before applying the $\pi$ theorem.

As applications, we first discuss falling motion under linear drag and clarify what the relation $F(\pi_1,\pi_2)=0$ supplied by the $\pi$ theorem means. We then treat the classical drag problem in a viscous fluid and compare the present framework with the usual elimination of redundant variables. Finally, we apply the method to a system of stock-market indicators. These indicators combine quantities such as price per share, earnings per share, book value per share, and dividends to describe the state of a firm or stock; examples include PER, PBR, ROE, and PR~\cite{cochrane2005}. Although they are usually introduced one by one through financial or accounting meanings, once appropriate dimensions are assigned they can be viewed as a single constrained system for dimensional analysis.

Two points should be emphasized. First, the drag example can also be treated by the conventional route, namely by eliminating the kinematic viscosity first and then applying the ordinary $\pi$ theorem. It is included as a benchmark example showing how the present method reproduces the standard result while keeping the constraint explicit. Second, treating stock-market indicators as an object of dimensional analysis appears to be new. The motivation is related to a common statement that ``stock-market indicators are dimensionless.'' This statement often confuses independence of firm size with dimensionlessness. In physical language, it is similar to saying that density is dimensionless because it is intensive.

The paper is organized as follows. In Sec.~\ref{sec:buckingham}, we review Buckingham's $\pi$ theorem in logarithmic variables and illustrate its meaning with falling motion under linear drag. In Sec.~\ref{sec:constraints}, we extend the formulation to constrained systems and derive both the effective number of independent dimensionless quantities and the redundancy elimination procedure. In Sec.~\ref{sec:drag}, we demonstrate the method using the drag problem in a viscous fluid. In Sec.~\ref{sec:stock}, we apply the framework to stock-market indicators. Section~\ref{sec:conclusion} gives conclusions and open directions.

\section{Review of Buckingham's \texorpdfstring{$\pi$}{pi} theorem}
\label{sec:buckingham}

\subsection{Dimension matrix and logarithmic variables}

We assume that there are $m$ formal dimensions $\DD_i$ $(1\leq i\leq m)$ and $n$ quantities $x_j$ $(1\leq j\leq n)$ in the system. The dimension of each quantity is written as
\eq{
\llb x_j \rrb
=
\prod_{i=1}^m \DD_i^{a_{ij}}
\qquad
(1\leq j\leq n).
\label{eq:dim-dependence}
}
The matrix $A\coloneqq [a_{ij}]\in {\mathbb R}^{m\times n}$ is called the dimension matrix.

In what follows, we assume $x_j>0$ and write ${\bm x}=(x_1,\dots,x_n)\in {\mathbb R}_{>0}^n$, because logarithmic variables will be used. This assumption is natural in many applications where the quantities are positive. If a component can cross zero, the logarithmic formulation cannot be applied directly and requires a separate treatment.

For an exponent vector ${\bm v}=(v_1,\dots,v_n)\in {\mathbb R}^n$, define the monomial
\eq{
{\bm x}^{\bm v}
\coloneqq
\prod_{j=1}^n x_j^{v_j}.
\label{eq:monomial}
}
Introducing the logarithmic variable
${\bm y}\coloneqq \ln {\bm x}=(\ln x_1,\dots,\ln x_n)\in {\mathbb R}^n$, we have
\eq{
\ln {\bm x}^{\bm v}
=
\sum_{j=1}^n v_j\ln x_j
=
\langle {\bm v},{\bm y}\rangle,
\label{eq:log-monomial}
}
where $\langle \ ,\ \rangle$ denotes the standard inner product. Therefore, under a change ${\bm y}\to {\bm y}+\delta{\bm y}$, the monomial changes according to
\eq{
\delta\ln {\bm x}^{\bm v}
=
\langle {\bm v},\delta{\bm y}\rangle.
\label{eq:delta-log-monomial}
}

\subsection{Dimensionless quantities and the \texorpdfstring{$\pi$}{pi} theorem}

We describe a change of units as a scaling of the formal dimensions:
\eq{
\DD_i\to e^{\lambda_i}\DD_i,
\qquad
\lambda_i\in {\mathbb R}
\qquad
(1\leq i\leq m).
\label{eq:dimension-scaling}
}
Here $\DD_i$ represents the scaling induced on the numerical value of a quantity by the change of unit. For example, changing the unit of mass from kg to g multiplies the numerical value of the same physical mass by $10^3$, which is represented by $\MM\to 10^3\MM$.

With this convention, Eq.~\eqref{eq:dim-dependence} implies that a change of units induces a translation in logarithmic variables:
\eq{
{\bm y}\to {\bm y}+A^\top{\bm\lambda},
\label{eq:unit-translation}
}
where ${\bm\lambda}=(\lambda_1,\dots,\lambda_m)$ and $\top$ denotes the transpose.

We now derive the condition under which ${\bm x}^{\bm e}$ is dimensionless. A dimensionless quantity is invariant under arbitrary changes of units. Thus we require
\eq{
\delta\ln {\bm x}^{\bm e}
=
\langle {\bm e},A^\top{\bm\lambda}\rangle
=
\langle A{\bm e},{\bm\lambda}\rangle
=0
\label{eq:delta-log-pi}
}
for all ${\bm\lambda}$. This holds if and only if ${\bm e}\in\ker A$. Therefore the number $d$ of independent dimensionless quantities is
\eq{
d\coloneqq \dim\ker A=n-\rank A.
\label{eq:buckingham1}
}

If $\{{\bm e}_1,\dots,{\bm e}_d\}$ is a basis of $\ker A$, Buckingham's $\pi$ theorem states that any physical relation in the system can be written as
\eq{
F(\pi_1,\pi_2,\dots,\pi_d)=0,
\qquad
\pi_k\coloneqq {\bm x}^{{\bm e}_k}
\qquad
(1\leq k\leq d),
\label{eq:buckingham2}
}
where $F:{\mathbb R}^d\to{\mathbb R}$ is an appropriate function.

\subsection{Falling motion under linear drag}

To clarify what the $\pi$ theorem provides, consider a body falling vertically under gravity $mg$ and a drag force $-kv$ proportional to the velocity. The body is released from rest, and we ask how the velocity $v$ is determined as a function of time $t$. Here $m$ is the mass, $g$ is the gravitational acceleration, and $k$ is the drag coefficient.

Assume first that
\eq{
v=v(t,m,g,k).
\label{eq:falling-functional-form}
}
Let the basic dimensions be $\DD_1=\MM$ for mass, $\DD_2=\LL$ for length, and $\DD_3=\TT$ for time. Thus $m=3$ in the notation of Sec.~\ref{sec:buckingham}. The variables are $v,t,m,g,k$, so $n=5$. Their dimensions are
\eq{
\llb v\rrb=\LL\TT^{-1},\qquad
\llb t\rrb=\TT,\qquad
\llb m\rrb=\MM,\qquad
\llb g\rrb=\LL\TT^{-2},\qquad
\llb k\rrb=\MM\TT^{-1}.
\label{eq:falling-dimensions}
}
With $(x_1,x_2,x_3,x_4,x_5)=(v,t,m,g,k)$, the dimension matrix is
\eq{
A=
\begin{bmatrix}
0&0&1&0&1\\
1&0&0&1&0\\
-1&1&0&-2&-1
\end{bmatrix}.
\label{eq:Afalling}
}
Since $\rank A=3$, Eq.~\eqref{eq:buckingham1} gives $d=5-3=2$. A possible basis of $\ker A$ is
\eq{
{\bm e}_1=(1,0,-1,-1,1),
\qquad
{\bm e}_2=(0,1,-1,0,1).
\label{eq:falling-kernel-basis}
}
The corresponding dimensionless quantities are
\eq{
\pi_1=\frac{vk}{mg},
\qquad
\pi_2=\frac{kt}{m}.
\label{eq:falling-pi}
}
Therefore the $\pi$ theorem asserts that the relation in this problem has the form
\eq{
F\!\left(\frac{vk}{mg},\frac{kt}{m}\right)=0.
\label{eq:fall-pi}
}

The important point is that the $\pi$ theorem supplies the form of the relation, but not the function $F$ itself. Solving Newton's equation
\eq{
m\frac{dv}{dt}=mg-kv,
\qquad
v(0)=0
\label{eq:fall-eq}
}
gives
\eq{
v(t)=\frac{mg}{k}\left(1-e^{-kt/m}\right).
\label{eq:fall-solution}
}
Thus Eq.~\eqref{eq:fall-pi} is concretely given by
\eq{
\frac{vk}{mg}=1-e^{-kt/m},
\label{eq:fall-pi-relation}
}
or equivalently
\eq{
F(\pi_1,\pi_2)
=
\pi_1-(1-e^{-\pi_2})
=0.
\label{eq:fall-F}
}
This example shows that the $\pi$ theorem identifies which dimensionless quantities should be considered, while equations, theory, experiments, or data are still needed to determine the relation among them. The interpretation of a relation such as $F(\mathrm{PBR},\mathrm{PR})=0$ in Sec.~\ref{sec:stock} is exactly of this kind.

\subsection{Orthogonal decomposition of logarithmic-variable space}

Before generalizing to constrained systems, we introduce a useful decomposition of the space of logarithmic variables.

By Eq.~\eqref{eq:delta-log-pi}, any dimensionless quantity is unchanged under variations of the form $\delta{\bm y}=A^\top{\bm\lambda}$, namely variations in $\im A^\top$. We call $\im A^\top$ the scaling direction and decompose the space of variations into this direction and its orthogonal complement:
\eq{
{\mathbb R}^n
=
\im A^\top\oplus\ker A.
\label{eq:decomp1}
}
The equality $(\im A^\top)^\perp=\ker A$ follows from the fundamental theorem of linear algebra~\cite{Strang}. We call the second component in Eq.~\eqref{eq:decomp1} the dimensionless direction, since motion along it changes dimensionless combinations and does not correspond to a scaling transformation.

Let
\eq{
E\coloneqq
[{\bm e}_1\ {\bm e}_2\ \dots\ {\bm e}_d]
\in {\mathbb R}^{n\times d}
\label{eq:E-def}
}
be the matrix whose columns form a basis of $\ker A$. Then an arbitrary infinitesimal variation can be written as
\eq{
\delta{\bm y}
=
A^\top{\bm\lambda}+E{\bm\eta}
\qquad
({\bm\lambda}\in{\mathbb R}^m,\ {\bm\eta}\in{\mathbb R}^d).
\label{eq:decomp2}
}
This decomposition is the basis of the following discussion.

\section{Extension to systems with constraints}
\label{sec:constraints}

\subsection{Constraint manifold and scale invariance}

Assume that the constraints are written as
${\bm\phi}({\bm x})={\bm 0}_{\ell}$ by a map
${\bm\phi}:{\mathbb R}^n\to{\mathbb R}^{\ell}$, where ${\bm 0}_{\ell}$ is the $\ell$-dimensional zero vector. We also assume that, in logarithmic variables, these constraints can be written equivalently as
\eq{
{\bm\psi}({\bm y})={\bm 0}_{\ell},
\label{eq:psi-constraint}
}
where ${\bm\psi}:{\mathbb R}^n\to{\mathbb R}^{\ell}$.

The constraint manifold is defined by the level set
\eq{
{\cal M}
\coloneqq
{\bm\psi}^{-1}({\bm 0}_{\ell}).
\label{eq:constraint-manifold}
}
At a point ${\bm y}\in{\cal M}$, the Jacobian matrix is
\eq{
J({\bm y})
\coloneqq
D{\bm\psi}({\bm y})
=
\left[
\frac{\pd\psi_k}{\pd y_j}
\right]_{1\leq k\leq \ell,\ 1\leq j\leq n}
\in{\mathbb R}^{\ell\times n}.
\label{eq:J-def}
}
If $J=D{\bm\psi}$ has locally constant rank on ${\cal M}$ near ${\bm y}$, then the constant rank theorem implies that ${\cal M}$ is a submanifold near ${\bm y}$~\cite{Lee2013}, and
\eq{
T_{\bm y}{\cal M}=\ker J({\bm y}),
\qquad
\dim T_{\bm y}{\cal M}=n-\rank J.
\label{eq:tangent-space}
}
The fundamental theorem of linear algebra also gives
\eq{
{\mathbb R}^n=\ker J\oplus\im J^\top.
\label{eq:J-decomp}
}

We now introduce scale-invariant constraints. A constraint is said to be scale invariant if, for every ${\bm\lambda}\in{\mathbb R}^m$, the scaling direction $A^\top{\bm\lambda}$ belongs to the tangent space of the constraint manifold. In other words,
\eq{
JA^\top{\bm\lambda}
=
{\bm 0}_{\ell}
\label{eq:scale-invariant-condition}
}
holds. Thus, for scale-invariant constraints,
\eq{
\im A^\top\subseteq\ker J
\qquad
(\text{scale-invariant constraints}).
\label{eq:imA^TinKerJ}
}

\subsection{Effective number of free dimensionless quantities}

Linearizing the constraints around ${\bm y}$, an admissible infinitesimal change $\delta{\bm y}$ must satisfy
\eq{
J\delta{\bm y}={\bm 0}_{\ell}.
\label{eq:Jy=0}
}
By Eq.~\eqref{eq:decomp2}, any infinitesimal variation decomposes as
$\delta{\bm y}=A^\top{\bm\lambda}+E{\bm\eta}$. The scaling part $A^\top{\bm\lambda}$ does not change dimensionless quantities. Therefore the dimensionless degrees of freedom are carried only by $E{\bm\eta}$, and it is sufficient to impose Eq.~\eqref{eq:Jy=0} on $\delta{\bm y}=E{\bm\eta}$. This gives
\eq{
JE{\bm\eta}={\bm 0}_{\ell}.
\label{eq:JEeta=0}
}
Thus the effective number of independent dimensionless quantities at ${\bm y}$ is defined by
\eq{
d_{\rm eff}
\coloneqq
\dim\ker JE.
\label{eq:deff1}
}

\begin{figure}[t]
\centering
\includegraphics[width=0.52\linewidth]{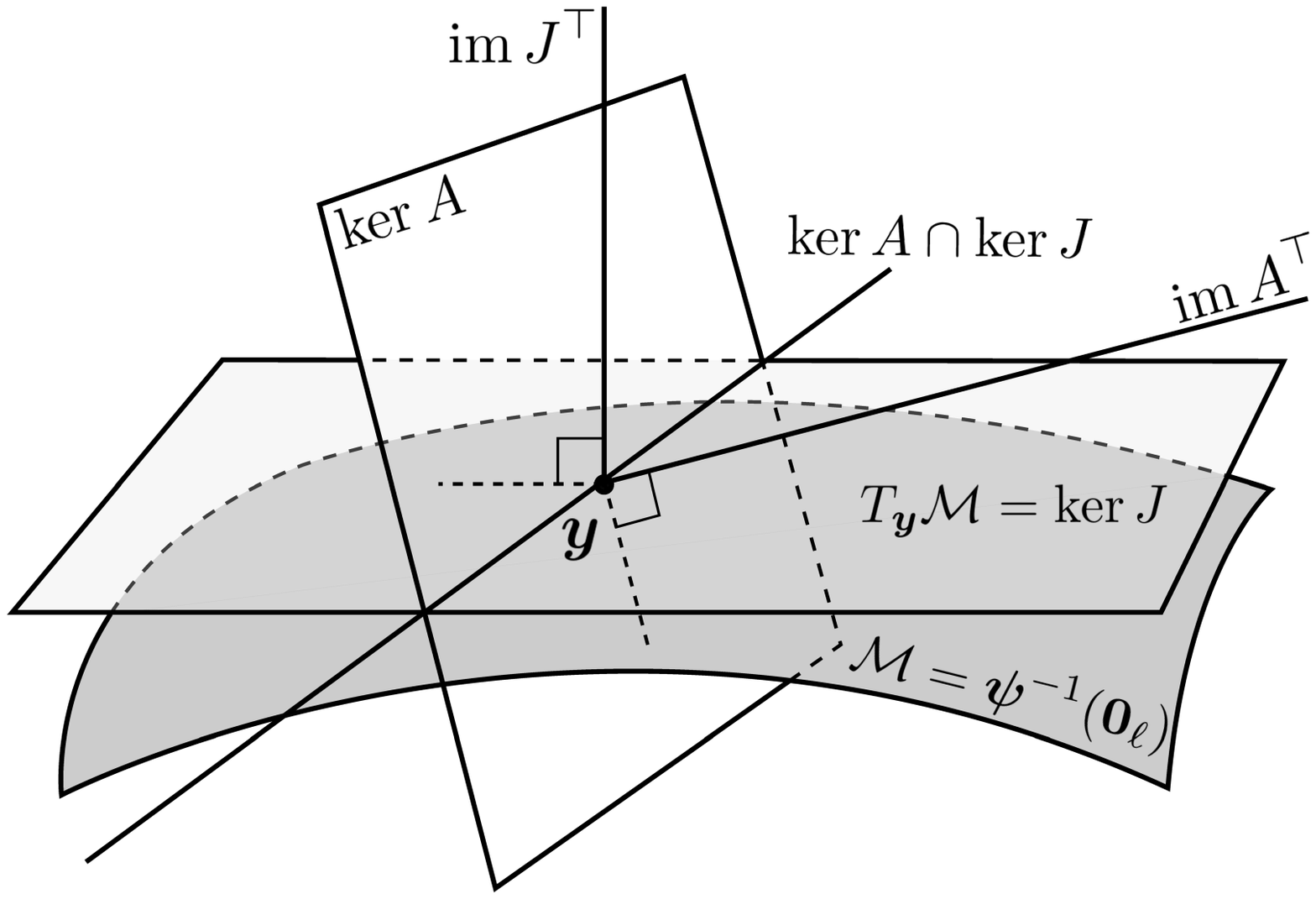}
\caption{A schematic picture of the space of logarithmic variables ${\bm y}=\ln{\bm x}$. The constraint manifold is ${\cal M}={\bm\psi}^{-1}({\bm 0}_{\ell})$, and its tangent space is $T_{\bm y}{\cal M}=\ker J$. The number $d_{\rm eff}$ of admissible dimensionless quantities is the dimension of $\ker A\cap\ker J$. For scale-invariant constraints, $\im A^\top\subseteq\ker J$.}
\label{fig:kerAkerJ}
\end{figure}

The matrix $E$ gives a natural identification between $\ker JE$ and $\ker A\cap\ker J$. Indeed,
\eq{
\ker JE
&=
\{
{\bm\eta}\in{\mathbb R}^d:
JE{\bm\eta}={\bm 0}_{\ell}
\}
\nn\\
&\cong
\{
\delta{\bm y}\in{\mathbb R}^n:
J\delta{\bm y}={\bm 0}_{\ell},
\ \delta{\bm y}\in\ker A
\}
\nn\\
&=
\ker A\cap\ker J.
\label{eq:kerJE-kerAkerJ}
}
Therefore
\eq{
d_{\rm eff}
=
\dim(\ker A\cap\ker J).
\label{eq:deff2}
}
This means that the number of admissible dimensionless quantities is the dimension of the intersection between the dimensionless direction $\ker A$ and the tangent direction $\ker J$ allowed by the constraint, as shown schematically in Fig.~\ref{fig:kerAkerJ}.

Equation~\eqref{eq:deff2} and the rank-nullity theorem also give the direct computational formula
\eq{
d_{\rm eff}
=
n-\rank
\begin{bmatrix}
A\\
J
\end{bmatrix}.
\label{eq:deff3}
}

Using the Grassmann dimension formula, we can also write
\eq{
d_{\rm eff}
&=
\dim\ker A+\dim\ker J-\dim(\ker A+\ker J)
\nn\\
&=
n-\rank A-\rank J
+
\dim(\im A^\top\cap\im J^\top).
\label{eq:deff4-pre}
}
If the constraints are scale invariant, Eq.~\eqref{eq:imA^TinKerJ} gives
$\im A^\top\subseteq\ker J=(\im J^\top)^\perp$, and hence
$\im A^\top\cap\im J^\top=\{{\bm 0}_n\}$. Therefore
\eq{
d_{\rm eff}
=
n-\rank A-\rank J
\qquad
(\text{scale-invariant constraints}).
\label{eq:deff4}
}

\subsection{Mechanical elimination of redundant dimensionless quantities}

The previous subsection gives the number of independent admissible dimensionless quantities. In applications, however, one often first constructs a basis $\{{\bm e}_1,\dots,{\bm e}_d\}$ of $\ker A$ and the corresponding candidate dimensionless quantities $\{\pi_1,\dots,\pi_d\}$, and then asks which of these candidates are redundant under the constraints.

Assume in this subsection that the constraints are scale invariant. Write
$J^\top=[{\bm j}_1\ {\bm j}_2\ \dots\ {\bm j}_{\ell}]$. From $JA^\top=O_{\ell\times m}$, we have
\eq{
A{\bm j}_k={\bm 0}_m
\qquad
(1\leq k\leq \ell).
\label{eq:jk-in-kerA}
}
Thus each ${\bm j}_k$ belongs to $\ker A$ and can be written as a linear combination of $\{{\bm e}_1,\dots,{\bm e}_d\}$. Equivalently, there exists a matrix $C\in{\mathbb R}^{\ell\times d}$ such that
\eq{
J=CE^\top.
\label{eq:JCE}
}
This matrix is uniquely determined by
\eq{
C=JE(E^\top E)^{-1}.
\label{eq:CJEEE}
}

Since $\pi_k={\bm x}^{{\bm e}_k}$, the vector
$\ln{\bm\pi}\coloneqq(\ln\pi_1,\dots,\ln\pi_d)$ is
\eq{
\ln{\bm\pi}
=
E^\top{\bm y}.
\label{eq:lnpi}
}
Substituting Eq.~\eqref{eq:JCE} into the infinitesimal constraint $J\delta{\bm y}={\bm 0}_{\ell}$ and using Eq.~\eqref{eq:lnpi}, we obtain
\eq{
C\delta\ln{\bm\pi}
=
{\bm 0}_{\ell}.
\label{eq:Cdeltapi}
}
When $C$ is constant, this relation integrates to
\eq{
C\ln{\bm\pi}
=
{\bm c},
\label{eq:Cpi}
}
where ${\bm c}\in{\mathbb R}^{\ell}$ is a constant vector determined by the constraint manifold.

Thus the linear relations among candidate dimensionless quantities are encoded by $C$. In particular, $\rank C$ is the number of redundant directions among the $d$ candidates, and
\eq{
d_{\rm eff}=d-\rank C.
\label{eq:deff-C}
}
Since Eqs.~\eqref{eq:JCE} and \eqref{eq:CJEEE} imply $\rank C=\rank J$, this agrees with Eq.~\eqref{eq:deff4}.

The procedure is summarized as follows.
\begin{enumerate}
\item Construct the dimension matrix $A$, choose a basis $\{{\bm e}_1,\dots,{\bm e}_d\}$ of $\ker A$, and form $E=[{\bm e}_1\ {\bm e}_2\ \dots\ {\bm e}_d]$.
\item Compute the Jacobian matrix $J$ of the constraints and verify the scale-invariance condition $JA^\top=O_{\ell\times m}$.
\item Compute $C$ from $J=CE^\top$ or, equivalently, from $C=JE(E^\top E)^{-1}$.
\item Reduce $C$ to row echelon form. Pivot columns correspond to variables determined by free variables, while non-pivot columns give one systematic choice of independent dimensionless quantities.
\end{enumerate}
In this way, redundant dimensionless quantities can be removed by elementary linear algebra once the basis of $\ker A$ and the Jacobian matrix $J$ are given.

\section{Illustration by the drag problem}
\label{sec:drag}

We now illustrate the formulation using the classical problem of drag in a viscous fluid~\cite{White2016}. Let $F_D$ be the drag force acting on a body with characteristic length $L$ and velocity $U$ in a fluid of density $\rho$, viscosity $\mu$, and kinematic viscosity $\nu$. Although the defining relation
\eq{
\nu\coloneqq\frac{\mu}{\rho}
\label{eq:nudef}
}
holds, we deliberately include $\nu$ as an independent variable and impose Eq.~\eqref{eq:nudef} later as a constraint. This makes it clear how redundant variables are treated in the present formulation.

For readers familiar with fluid mechanics, this example may look unnecessarily indirect. Indeed, in the conventional treatment, one first uses Eq.~\eqref{eq:nudef} to eliminate $\nu$, applies the ordinary $\pi$ theorem to the five variables $(F_D,\rho,U,L,\mu)$, and obtains the drag coefficient and the Reynolds number. We start instead from six variables in order to show that the same procedure applies even when such prior elimination is not convenient.

Let the basic dimensions be $\DD_1=\MM$ for mass, $\DD_2=\LL$ for length, and $\DD_3=\TT$ for time. With
\eq{
(x_1,x_2,x_3,x_4,x_5,x_6)
=(F_D,\rho,U,L,\mu,\nu),
\qquad
n=6,
\label{eq:drag-variables}
}
the dimension matrix is
\eq{
A=
\begin{bmatrix}
1 & 1 & 0 & 0 & 1 & 0\\
1 & -3 & 1 & 1 & -1 & 2\\
-2 & 0 & -1 & 0 & -1 & -1
\end{bmatrix}.
\label{eq:Adrag}
}
For example, the first column represents $\llb F_D\rrb=\MM^1\LL^1\TT^{-2}$, and the second column represents $\llb\rho\rrb=\MM^1\LL^{-3}\TT^0$. Since $\rank A=3$, the number of candidate dimensionless quantities before imposing the constraint is $d=6-3=3$.

One basis of $\ker A$ is
\eq{
{\bm e}_1=(1,-1,-2,-2,0,0),
\qquad
{\bm e}_2=(0,1,1,1,-1,0),
\qquad
{\bm e}_3=(0,0,1,1,0,-1).
\label{eq:drag-kernel-basis}
}
The corresponding dimensionless quantities are
\eq{
\pi_1=\frac{F_D}{\rho U^2L^2},
\qquad
\pi_2=\frac{\rho UL}{\mu},
\qquad
\pi_3=\frac{UL}{\nu}.
\label{eq:drag-pi}
}
In fluid mechanics, $\pi_1$ is proportional to the drag coefficient $C_D$, and $\pi_2$ is the Reynolds number $Re$.

The constraint \eqref{eq:nudef} is written in logarithmic variables as
\eq{
\psi({\bm y})=y_2-y_5+y_6=0.
\label{eq:drag-psi}
}
Thus
\eq{
J=D\psi({\bm y})
=
\begin{bmatrix}
0&1&0&0&-1&1
\end{bmatrix},
\label{eq:Jdrag}
}
and $\rank J=1$. From Eqs.~\eqref{eq:Adrag} and \eqref{eq:Jdrag}, one verifies $JA^\top=O_{1\times3}$, so the constraint is scale invariant. Therefore
\eq{
d_{\rm eff}
=
n-\rank A-\rank J
=2.
\label{eq:drag-deff}
}

Let $E=[{\bm e}_1\ {\bm e}_2\ {\bm e}_3]$. Computing $C$ from Eq.~\eqref{eq:JCE} or Eq.~\eqref{eq:CJEEE} gives
\eq{
C=
\begin{bmatrix}
0&1&-1
\end{bmatrix}.
\label{eq:Cdrag}
}
This matrix is already in row echelon form, with the second column as a pivot column. Therefore
\eq{
\delta\ln\pi_2-\delta\ln\pi_3=0
\label{eq:drag-delta-pi}
}
and integration gives
\eq{
\frac{\pi_2}{\pi_3}
=
\mathrm{const.}
\label{eq:drag-pi-ratio}
}
In fact this constant is $1$, and one of $\pi_2$ and $\pi_3$ is redundant.

A convenient independent set is therefore $\pi_1,\pi_2$. The drag relation can be written as $F(\pi_1,\pi_2)=0$, and locally as
\eq{
C_D=f(Re).
\label{eq:drag-law}
}
This is the classical statement that the drag coefficient is organized as a function of the Reynolds number. The conventional treatment first uses $\nu=\mu/\rho$ and then applies the $\pi$ theorem. In the present treatment, one keeps the constraint, constructs candidate $\pi$ groups, and removes redundancy at the end by the matrix $C$.

\section{Application to stock-market indicators}
\label{sec:stock}

We now apply the framework to a system of stock-market indicators. Such indicators are used in equity markets and firm valuation to express relations among quantities such as price per share, earnings per share, book value per share, and dividends. In financial practice, the indicators are often interpreted individually. From the viewpoint of dimensional analysis, however, they form a constrained system consisting of dimensional quantities and defining equations. A short glossary of the symbols used below is given in Appendix~\ref{sec:finance-glossary}.

The purpose here is not to discuss the investment interpretation of each indicator, but to see how the defining equations tie them together and reduce the degrees of freedom.

\subsection{Variables and dimensionless quantities}

Define the variable vector ${\bm x}=(x_1,x_2,\dots,x_{10})$ by
\eq{
\begin{alignedat}{2}
&x_1=\mathrm{PPS}=\text{Price per Share},\quad&
&x_2=\mathrm{EPS}=\text{Earnings per Share},\\
&x_3=\mathrm{BPS}=\text{Book Value per Share},\quad&
&x_4=\mathrm{DPS}=\text{Dividend per Share},\\
&x_5=\mathrm{PER}=\text{Price-to-Earnings Ratio},\quad&
&x_6=\mathrm{PBR}=\text{Price-to-Book Ratio},\\
&x_7=\mathrm{EY}=\text{Earnings Yield},\quad&
&x_8=\mathrm{ROE}=\text{Return on Equity},\\
&x_9=\mathrm{DY}=\text{Dividend Yield},\quad&
&x_{10}=\mathrm{PR}=\text{Payout Ratio}.
\end{alignedat}
\label{eq:stock-variables}
}
We use currency $\DD_1=\CC$, shares $\DD_2=\SSS$, and time $\DD_3=\TT$ as basic dimensions. The dimensions of the variables are
\eq{
\begin{alignedat}{3}
\llb \mathrm{PPS}\rrb&=\CC\SSS^{-1},\quad&
\llb \mathrm{EPS}\rrb&=\CC\SSS^{-1}\TT^{-1},\quad&
\llb \mathrm{BPS}\rrb&=\CC\SSS^{-1},\\
\llb \mathrm{DPS}\rrb&=\CC\SSS^{-1}\TT^{-1},\quad&
\llb \mathrm{PER}\rrb&=\TT,\quad&
\llb \mathrm{PBR}\rrb&=1,\\
\llb \mathrm{EY}\rrb&=\TT^{-1},\quad&
\llb \mathrm{ROE}\rrb&=\TT^{-1},\quad&
\llb \mathrm{DY}\rrb&=\TT^{-1},\\
&&\llb \mathrm{PR}\rrb&=1.
\end{alignedat}
\label{eq:stock-dimensions}
}
The dimension matrix is therefore
\eq{
A=
\begin{bmatrix}
1&1&1&1&0&0&0&0&0&0\\
-1&-1&-1&-1&0&0&0&0&0&0\\
0&-1&0&-1&1&0&-1&-1&-1&0
\end{bmatrix}.
\label{eq:Astock}
}
The second row is $-1$ times the first row, so $\rank A=2$ and
\eq{
d=n-\rank A=8.
\label{eq:stock-d}
}

Choose the following basis of $\ker A$:
\eq{
\begin{alignedat}{2}
{\bm e}_1&=(1,0,0,-1,-1,0,0,0,0,0),\quad&
{\bm e}_2&=(0,1,0,-1,0,0,0,0,0,0),\\
{\bm e}_3&=(0,0,1,-1,-1,0,0,0,0,0),\quad&
{\bm e}_4&=(0,0,0,0,1,0,1,0,0,0),\\
{\bm e}_5&=(0,0,0,0,1,0,0,1,0,0),\quad&
{\bm e}_6&=(0,0,0,0,1,0,0,0,1,0),\\
{\bm e}_7&=(0,0,0,0,0,1,0,0,0,0),\quad&
{\bm e}_8&=(0,0,0,0,0,0,0,0,0,1).
\end{alignedat}
\label{eq:stock-kernel-basis}
}
From $\pi_k\coloneqq{\bm x}^{{\bm e}_k}$, we obtain
\eq{
\begin{alignedat}{4}
\pi_1&=\frac{\mathrm{PPS}}{\mathrm{DPS}\cdot\mathrm{PER}},\quad&
\pi_2&=\frac{\mathrm{EPS}}{\mathrm{DPS}},\quad&
\pi_3&=\frac{\mathrm{BPS}}{\mathrm{DPS}\cdot\mathrm{PER}},\quad&
\pi_4&=\mathrm{PER}\cdot\mathrm{EY},\\
\pi_5&=\mathrm{PER}\cdot\mathrm{ROE},\quad&
\pi_6&=\mathrm{PER}\cdot\mathrm{DY},\quad&
\pi_7&=\mathrm{PBR},\quad&
\pi_8&=\mathrm{PR}.
\end{alignedat}
\label{eq:stock-pi}
}

\subsection{Constraints and reduction}

We impose the definitions of the stock-market indicators as constraints:
\eq{
\begin{alignedat}{2}
\mathrm{PER}&\coloneqq\frac{\mathrm{PPS}}{\mathrm{EPS}},\quad&
\mathrm{PBR}&\coloneqq\frac{\mathrm{PPS}}{\mathrm{BPS}},\\
\mathrm{EY}&\coloneqq\frac{\mathrm{EPS}}{\mathrm{PPS}},\quad&
\mathrm{ROE}&\coloneqq\frac{\mathrm{EPS}}{\mathrm{BPS}},\\
\mathrm{DY}&\coloneqq\frac{\mathrm{DPS}}{\mathrm{PPS}},\quad&
\mathrm{PR}&\coloneqq\frac{\mathrm{DPS}}{\mathrm{EPS}}.
\end{alignedat}
\label{eq:stock-definitions}
}
In logarithmic variables ${\bm y}=\ln{\bm x}$, these become the linear constraints
\eq{
\begin{alignedat}{2}
\psi_1({\bm y})&=y_5-y_1+y_2=0,\quad&
\psi_2({\bm y})&=y_6-y_1+y_3=0,\\
\psi_3({\bm y})&=y_7-y_2+y_1=0,\quad&
\psi_4({\bm y})&=y_8-y_2+y_3=0,\\
\psi_5({\bm y})&=y_9-y_4+y_1=0,\quad&
\psi_6({\bm y})&=y_{10}-y_4+y_2=0.
\end{alignedat}
\label{eq:stock-psi}
}
The Jacobian matrix is
\eq{
J=
\begin{bmatrix}
-1&1&0&0&1&0&0&0&0&0\\
-1&0&1&0&0&1&0&0&0&0\\
1&-1&0&0&0&0&1&0&0&0\\
0&-1&1&0&0&0&0&1&0&0\\
1&0&0&-1&0&0&0&0&1&0\\
0&1&0&-1&0&0&0&0&0&1
\end{bmatrix}.
\label{eq:Jstock}
}
This is constant because the constraints are linear in logarithmic variables.

Computing $C$ from Eq.~\eqref{eq:CJEEE}, we find
\eq{
C=
\begin{bmatrix}
-1&1&0&0&0&0&0&0\\
-1&0&1&0&0&0&1&0\\
1&-1&0&1&0&0&0&0\\
0&-1&1&0&1&0&0&0\\
1&0&0&0&0&1&0&0\\
0&1&0&0&0&0&0&1
\end{bmatrix}.
\label{eq:Cstock}
}
A direct computation gives $\rank C=6$. Equation~\eqref{eq:Cpi} can be rewritten as the following algebraic relations among the $\pi_k$:
\eq{
\begin{aligned}
\pi_4&=\mathrm{const.},\qquad&
\frac{\pi_5}{\pi_7}&=\mathrm{const.},\qquad&
\frac{\pi_6}{\pi_8}&=\mathrm{const.},\\
\pi_1\pi_8&=\mathrm{const.},\qquad&
\pi_2\pi_8&=\mathrm{const.},\qquad&
\pi_3\pi_7\pi_8&=\mathrm{const.}.
\end{aligned}
\label{eq:stock-pi-relations}
}
The first three relations correspond to the familiar identities
\eq{
\mathrm{EY}=\frac{1}{\mathrm{PER}},
\qquad
\mathrm{ROE}=\frac{\mathrm{PBR}}{\mathrm{PER}},
\qquad
\mathrm{DY}=\frac{\mathrm{PR}}{\mathrm{PER}}.
\label{eq:stock-identities}
}

In finance, these relations are usually introduced as separate definitions or accounting identities. In the present formulation, they are derived from the same linear-algebraic structure governed by the single matrix $C$. Thus the identities are not independent facts, but different representations of the low-dimensional structure created by the definitions.

The number of independent dimensionless quantities is
\eq{
d_{\rm eff}=d-\rank C=2.
\label{eq:stock-deff}
}
The choice of two quantities is not unique; it corresponds to a change of basis in $\ker A$, as discussed in Appendix~\ref{sec:basis-tr}. If one wants to retain practically familiar variables, $\mathrm{PBR}$ and $\mathrm{PR}$ can be chosen as independent variables. In this sense, the seemingly diverse family of stock-market indicators reduces, under the defining equations alone, to two independent dimensionless degrees of freedom.

This viewpoint has a consequence beyond reorganizing well-known identities. Dimensional analysis says that any additional relation among the ten quantities, if it exists beyond the defining equations, can appear essentially only as a relation between two independent dimensionless quantities. For example, if $\mathrm{PBR}$ and $\mathrm{PR}$ are chosen as the independent quantities, an additional relation should have the form
\eq{
F(\mathrm{PBR},\mathrm{PR})=0.
\label{eq:stock-extra-relation}
}
This is the same type of statement as in the falling-body example, where the relation between $v$ and $t$ ultimately appeared as the relation $\pi_1=1-e^{-\pi_2}$ between two dimensionless quantities. Therefore, if a relation of the form \eqref{eq:stock-extra-relation} is found in market data, it means that the system of stock-market indicators is constrained not only by definitions but also by an additional regularity shared in the market.

\subsection{Preliminary examination with S\&P 500 data}

The preceding discussion does not assert that an additional empirical relation exists. It only specifies the pair of reduced variables in which such a relation should be sought. To see this concretely, we made a preliminary examination of the relation between PBR and PR for S\&P 500 constituents. The data were firm-level fundamentals obtained and stored from the EOD Historical Data (EODHD) fundamentals API as of March 2026~\cite{EODHD}. PBR was recomputed as price per share divided by book value per share, where price per share was obtained by dividing market capitalization by outstanding shares. PR was recomputed as dividend per share divided by earnings per share.

Consistently with the logarithmic formulation, only firms with finite and positive PBR and PR were used. More concretely, market capitalization, outstanding shares, book value per share, dividend per share, and earnings per share were required to be finite and positive. This left $N=360$ firms. As a sensitivity check, we also considered the sample obtained by excluding firms with payout ratio greater than one; that sample had $N=322$ firms.

As the simplest empirical model, assume
\eq{
\mathrm{PBR}=k\mathrm{PR}^{a},
\qquad
k>0.
\label{eq:sp500-power}
}
We estimated
\eq{
\ln(\mathrm{PBR})=b+a\ln(\mathrm{PR}),
\qquad
b=\ln k,
\label{eq:sp500-logreg}
}
by ordinary least squares. The results are summarized in Table~\ref{tab:sp500-reg}.

\begin{table}[htbp]
\centering
\caption{Preliminary regression of $\ln(\mathrm{PBR})=b+a\ln(\mathrm{PR})$ for the S\&P 500 data.}
\label{tab:sp500-reg}
\begin{tabular}{lrrrrr}
\hline
Sample & $N$ & $a$ & $k$ & $R^2$ & $p$-value \\
\hline
All sectors & 360 & $-0.171$ & $3.281$ & $0.023$ & $0.00399$ \\
All sectors, $\mathrm{PR}\leq 1$ & 322 & $-0.144$ & $3.377$ & $0.012$ & $0.0488$ \\
Communication Services & 17 & $-0.668$ & $1.569$ & $0.530$ & $0.000923$ \\
\hline
\end{tabular}
\end{table}

\begin{figure}[htbp]
\centering
\includegraphics[width=0.50\linewidth]{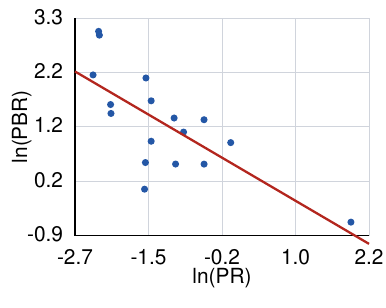}
\caption{Relation between $\ln(\mathrm{PR})$ and $\ln(\mathrm{PBR})$ in the Communication Services sector. The straight line is the least-squares regression line.}
\label{fig:communication-services}
\end{figure}

For all sectors combined, the slope in Eq.~\eqref{eq:sp500-logreg} is statistically different from zero, but $R^2$ is very small. Thus this single-year cross-sectional data set does not support the view that Eq.~\eqref{eq:sp500-power} is a strong market-wide law. On the other hand, the fit varies substantially by sector. In the Communication Services sector, a relatively clear negative correlation is visible, as shown in Fig.~\ref{fig:communication-services}. Since the sample size in that sector is only $N=17$, however, this result should be interpreted cautiously.

The point with $\mathrm{PR}>1$ in the lower right of Fig.~\ref{fig:communication-services} lies near the edge of the explanatory-variable range and is a high-leverage point for the regression coefficient. Removing it amounts to imposing $\mathrm{PR}\leq 1$ in the Communication Services sector. Even then the slope remains negative, $a=-0.764$, but $R^2$ decreases to $0.382$. Therefore the sector result should be read as an exploratory signal that motivates further analysis, rather than as a robust empirical law.

This preliminary examination illustrates that the present framework does not by itself produce empirical relations. Its role is to identify the reduced coordinates in which additional empirical structure should be sought after the defining constraints have been removed. The results above suggest that a strong power law is difficult to see when the S\&P 500 is treated as a single population, while some sector-dependent structure may be present. A serious financial-data application would require more careful statistical treatment of time dependence, sector classification, outliers, and firms with small earnings.

\section{Conclusion}
\label{sec:conclusion}

We formulated dimensional analysis for constrained systems as linear algebra in logarithmic variables. The effective number of independent dimensionless quantities is given by Eq.~\eqref{eq:deff2}, namely
\eq{
d_{\rm eff}
=
\dim(\ker A\cap\ker J),
\label{eq:conclusion-deff-intersection}
}
which is the dimension of the intersection between the dimensionless direction and the tangent direction compatible with the constraints. For computation, Eq.~\eqref{eq:deff3} gives
\eq{
d_{\rm eff}
=
n-\rank
\begin{bmatrix}
A\\
J
\end{bmatrix}.
\label{eq:conclusion-deff-rank}
}
If the constraints are scale invariant, Eq.~\eqref{eq:deff4} reduces this to
\eq{
d_{\rm eff}
=
n-\rank A-\rank J.
\label{eq:conclusion-deff-scale}
}

The main contribution is not only counting, but also a systematic method for eliminating redundant candidate dimensionless quantities. By writing the constraint matrix as $J=CE^\top$, as in Eq.~\eqref{eq:JCE}, dependencies among candidate dimensionless quantities are expressed by
\eq{
C\delta\ln{\bm\pi}
=
{\bm 0}_{\ell}.
\label{eq:conclusion-Cdelta}
}
Row reduction of $C$ then gives an independent set of dimensionless quantities. This replaces the often heuristic selection of $\pi$ groups by a procedure based on elementary linear algebra.

In the drag example, even when the kinematic viscosity is included as a redundant variable, the constraint $\nu=\mu/\rho$ in Eq.~\eqref{eq:nudef} implies $\pi_2/\pi_3=1$ and reduces the description to the drag coefficient and the Reynolds number. In the stock-market indicator example, relations among PER, PBR, EY, ROE, DY, and PR are all derived from the same linear-algebraic structure, and the defining equations reduce the system to two independent dimensionless degrees of freedom.

The framework is expected to be useful especially when constraints are complicated and explicit variable elimination is difficult. Examples include systems with many definitions or constitutive laws, or systems where constraints are given implicitly. The next step is to find examples in which this formulation is genuinely advantageous, and to characterize empirical or theoretical relations among the reduced invariants. In the stock-market example, if $\mathrm{PBR}$ and $\mathrm{PR}$ are chosen as independent quantities, an additional empirical relation would be written schematically as
\eq{
\mathrm{PBR}=f(\mathrm{PR}).
\label{eq:conclusion-stock-relation}
}
The preliminary S\&P 500 examination in this paper does not support a universal power law across all sectors, but it suggests that sector-dependent structures may be worth studying. More careful analysis of financial data, including time dependence, sector dependence, and outlier treatment, is left for future work. Extensions to cases where the rank of the Jacobian varies on the constraint manifold, and to constraints arising from data-driven models or more complex constitutive laws, are also important directions.

\section*{Acknowledgments}

The author thanks Yosuke Araya, Yoshiaki Shimazaki, Manato Haigo, and Kenta Hioki for useful advice and discussions related to this work. This work was supported by internal research funding from Akita Prefectural University.

\appendix

\section{Glossary of stock-market quantities and indicators}
\label{sec:finance-glossary}

For readers who are not familiar with financial practice, we summarize the symbols used in Sec.~\ref{sec:stock}. The explanations are intentionally brief and only provide the background needed for the dimensional-analysis example.

\noindent\textbf{PPS, price per share}: the market price of one share, directly observed in the stock market.

\noindent\textbf{EPS, earnings per share}: earnings attributable to one share over a fixed period; a per-share measure of profitability.

\noindent\textbf{BPS, book value per share}: book value of equity attributable to one share; a balance-sheet based measure of firm value per share.

\noindent\textbf{DPS, dividend per share}: dividends paid per share over a fixed period; a basic measure of cash returned to shareholders.

\noindent\textbf{PER, price-to-earnings ratio}: the number of years of earnings represented by the stock price; a common valuation multiple comparing price with earnings.

\noindent\textbf{PBR, price-to-book ratio}: the ratio of price per share to book value per share; a valuation multiple comparing market value with book value.

\noindent\textbf{EY, earnings yield}: the ratio of earnings to price; the reciprocal of PER.

\noindent\textbf{ROE, return on equity}: the ratio of earnings to book value; a measure of how efficiently equity is converted into earnings.

\noindent\textbf{DY, dividend yield}: the ratio of dividends to price; a common indicator for viewing a stock from the perspective of dividend income.

\noindent\textbf{PR, payout ratio}: the ratio of dividends to earnings; a measure of how much of earnings is paid out to shareholders.

In practice, PPS, EPS, BPS, and DPS are basic per-share quantities, while PER, PBR, EY, ROE, DY, and PR are ratios constructed from them. PER and PBR are valuation multiples, EY, ROE, and DY concern profitability or yield, and PR describes the allocation of earnings to dividends. Actual investment decisions require many other factors, including growth, leverage, sector characteristics, and accounting conventions. The point of Sec.~\ref{sec:stock} is only that these commonly listed indicators can be viewed as one constrained dimensional-analysis system.

\section{Basis transformation and invariance of the reduction}
\label{sec:basis-tr}

In the main text, candidate dimensionless quantities were constructed from a basis of $\ker A$, and redundant quantities were removed by the matrix $C$. The basis of $\ker A$ is not unique. This appendix records how the construction changes under a basis transformation and shows that the content of the reduction does not depend on the chosen basis.

\subsection{Transformation of the basis and constraint matrix}

Let $E=[{\bm e}_1\ {\bm e}_2\ \dots\ {\bm e}_d]$ be a basis matrix for $\ker A$, and let $M\in GL_d({\mathbb R})$ be invertible. Define a new basis by
\eq{
\tilde E=EM.
\label{eq:Etilde}
}
The corresponding dimensionless quantities satisfy
\eq{
\ln\tilde{\bm\pi}
=
M^\top\ln{\bm\pi}.
\label{eq:pitildeM}
}
Thus a basis transformation in $\ker A$ induces a linear transformation of logarithmic dimensionless variables.

The constraint matrix $C$ is defined by $J=CE^\top$. Under the change $E\to\tilde E=EM$, we have
\eq{
J
=
\tilde C\tilde E^\top
=
\tilde C M^\top E^\top,
\label{eq:J-Ctilde}
}
and hence
\eq{
\tilde C
=
C(M^\top)^{-1}.
\label{eq:Ctilde}
}
Therefore $C$ itself depends on the chosen basis, but its rank is invariant. In particular,
\eq{
d_{\rm eff}=d-\rank C
\label{eq:deff-basis-invariant}
}
does not depend on the basis.

The coordinates $\ln{\bm\pi}=E^\top{\bm y}$ represent coordinates along the dimensionless direction $\ker A$. Equation~\eqref{eq:pitildeM} is merely a coordinate transformation, and the admissible subspace defined by $C\delta\ln{\bm\pi}=0$ is the same geometric object written in different coordinates.

\subsection{Example for stock-market indicators}

For the stock-market example in Sec.~\ref{sec:stock}, consider another basis $\tilde E=EM$ with
\eq{
M=
\begin{bmatrix}
1&-2&0&0&0&0&0&0\\
0&1&0&0&0&2&0&0\\
0&0&1&0&0&0&1&0\\
0&0&0&1&0&-2&0&3\\
0&0&0&0&1&0&0&0\\
0&0&0&0&-1&1&0&0\\
-2&0&0&-2&0&0&1&0\\
0&0&0&0&0&0&0&1
\end{bmatrix}
\in GL_8({\mathbb R}).
\label{eq:stock-M}
}
Since $M$ is invertible, $\tilde E$ is also a basis of $\ker A$. For the corresponding dimensionless quantities $\tilde{\bm\pi}$, Eq.~\eqref{eq:pitildeM} gives
\eq{
\begin{alignedat}{4}
\tilde\pi_1&=\pi_1\pi_7^{-2},\quad&
\tilde\pi_2&=\pi_1^{-2}\pi_2,\quad&
\tilde\pi_3&=\pi_3,\quad&
\tilde\pi_4&=\pi_4\pi_7^{-2},\\
\tilde\pi_5&=\frac{\pi_5}{\pi_6},\quad&
\tilde\pi_6&=\pi_2^2\pi_4^{-2}\pi_6,\quad&
\tilde\pi_7&=\pi_3\pi_7,\quad&
\tilde\pi_8&=\pi_4^3\pi_8.
\end{alignedat}
\label{eq:stock-pitilde}
}
The transformed constraint matrix is given by Eq.~\eqref{eq:Ctilde}, and direct computation gives $\rank\tilde C=\rank C=6$. Therefore $d_{\rm eff}=2$ is unchanged. Row reduction may select different independent dimensionless quantities depending on the basis, but each choice is simply a different coordinate representation of the same two-dimensional admissible invariant space.

\end{document}